\documentclass[runningheads]{llncs}

\usepackage{graphicx}
\usepackage{tikz}
\usepackage{xcolor}

\usepackage{verbatim} % for comments

\usepackage{caption}
\usepackage{subcaption}

\usepackage{booktabs}
\usepackage{multirow}

\usepackage{hyperref}
\hypersetup{pdfborder=0 0 0}
\usepackage{cleveref}
\definecolor{verygoodgreen}{RGB}{17,115,22}
\definecolor{goodgreen}{RGB}{23,163,31}
\definecolor{slightlygoodgreen}{RGB}{181,230,29}
\definecolor{unclearyellow}{RGB}{248,197,61}
\definecolor{slightlybadred}{RGB}{233,125,16}
\definecolor{badred}{RGB}{177,2,2}
\definecolor{verybadred}{RGB}{136,0,21}

\newcommand{\impactverygood}{\tikz[baseline=(X.base)] \node[fill=verygoodgreen, rectangle, rounded corners=4pt, text=white] (X) {+3};}
\newcommand{\impactgood}{\tikz[baseline=(X.base)] \node[fill=goodgreen, rectangle, rounded corners=4pt, text=white] (X) {+2};}
\newcommand{\impactslightlygood}{\tikz[baseline=(X.base)] \node[fill=slightlygoodgreen, rectangle, rounded corners=4pt, text=black] (X) {+1};}
\newcommand{\impactnone}{\tikz[baseline=(X.base)] \node[fill=unclearyellow, rectangle, rounded corners=4pt, text=black] (X) {$\pm 0$};}
\newcommand{\impactslightlybad}{\tikz[baseline=(X.base)] \node[fill=slightlybadred, rectangle, rounded corners=4pt, text=black] (X) {-1};}
\newcommand{\impactbad}{\tikz[baseline=(X.base)] \node[fill=badred, rectangle, rounded corners=4pt, text=white] (X) {-2};}
\newcommand{\impactverybad}{\tikz[baseline=(X.base)] \node[fill=verybadred, rectangle, rounded corners=4pt, text=white] (X) {-3};}

\newcommand \copyrighttext {
  \footnotesize This preprint has not undergone peer review or any post-submission improvements or corrections. The Version of Record of this contribution is published in ``Requirements Engineering: Foundation for Software Quality'' (\url{https://doi.org/10.1007/978-3-031-57327-9}), and is available online at \url{https://doi.org/10.1007/978-3-031-57327-9_2}.
}

\newcommand\copyrightnotice{
    \begin{tikzpicture}[remember picture,overlay]
    \node[anchor=south,yshift=40pt] at (current page.south) {\fbox{\parbox{\dimexpr\textwidth-\fboxsep-\fboxrule\relax}{\copyrighttext}}};
    \end{tikzpicture}
}

\begin{document}

\title{Identifying relevant Factors of Requirements Quality: an industrial Case Study
%\thanks{This work was supported by the KKS foundation through the S.E.R.T. Research Profile project at Blekinge Institute of Technology.}
}
\titlerunning{Relevant Factors of Requirements Quality}

\author{Julian Frattini\inst{1}\orcidID{0000-0003-3995-6125}}

\authorrunning{Frattini, J.}

\institute{Blekinge Institute of Technology, Valhallavägen 1, 371 41 Karlskrona, Sweden \email{julian.frattini@bth.se}}

\maketitle
\copyrightnotice

\begin{abstract}
    \textbf{[Context and Motivation]}: % situate and motivate your research.
    The quality of requirements specifications impacts subsequent, dependent software engineering activities.
    Requirements quality defects like ambiguous statements can result in incomplete or wrong features and even lead to budget overrun or project failure.
    \textbf{[Problem]}: % formulate the specific question/problem addressed by the paper.
    Attempts at measuring the impact of requirements quality have been held back by the vast amount of interacting factors.
    Requirements quality research lacks an understanding of which factors are relevant in practice.
    \textbf{[Principal Ideas and Results]}: %summarize the ideas and results described in your paper. State, where appropriate, your research approach and methodology.
    We conduct a case study considering data from both interview transcripts and issue reports to identify relevant factors of requirements quality.
    The results include 17 factors and 11 interaction effects relevant to the case company.
    \textbf{[Contribution]}: %state the main contribution of your paper, by highlighting its added value (e.g., to theory, to practice). Also, state the limitations of your results.
    The results contribute empirical evidence that (1) strengthens existing requirements engineering theories and (2) advances industry-relevant requirements quality research.
    %While the result cannot be generalized, it serves as an indicator for similar industrial contexts.
\keywords{Requirements quality \and Case study \and Interview}
\end{abstract}

\section{Introduction}
\label{sec:intro}

Requirements specifications constitute the input to many subsequent software engineering activities and artifacts.
Requirements specifications are used to design architecture, develop code, or derive test cases.
Hence, the quality of requirements specifications impacts the software engineering process~\cite{mendez2017naming,wagner2019status}.

The requirements quality research domain aims to aid practitioners in understanding and managing the quality of their requirements specifications by detecting and removing requirements quality defects~\cite{montgomery2022empirical}. 
However, empirical evidence in this research domain remains scarce~\cite{frattini2022live}.
This hampers the adoption of research results in practice~\cite{femmer2018requirements,franch2020practitioners}.
Recent systematic studies of the requirements quality research domain have identified a lack of industrial relevance as a main factor holding the field back~\cite{montgomery2022empirical}.
%Recent systematic studies of the requirements quality research domain have identified a lack of industrial relevance~\cite{ivarsson2011method} as a main factor holding the field back~\cite{montgomery2022empirical}.

The aim of this study is to contribute empirical evidence about the impact of requirements quality defects by identifying factors of requirements quality that are relevant in practice.
To this end, an industrial case study at Ericsson has been conducted.
The gathered evidence both strengthens existing requirements engineering theories and steers future research efforts toward solving practically relevant problems.
While the results cannot be generalized due to the employed research method, we encourage replication in different industrial contexts with the disclosure of our replication package\footnote{Available at \url{https://doi.org/10.5281/zenodo.10149475}}.

The remainder of this article is structured as follows: \Cref{sec:background} introduces related work on requirements quality research. \Cref{sec:method} describes the applied method and \Cref{sec:result} reports the obtained results. These results are discussed in \Cref{sec:discussion} before concluding in \Cref{sec:conclusion}.

\section{Background}
\label{sec:background}

%\Cref{sec:background:rq} summarizes relevant aspects of requirements quality.
%\Cref{sec:background:rqt} presents previous work and the theoretical background used throughout this work.
%\Cref{sec:background:gap} outlines the remaining gaps which this work addresses.

\subsection{Requirements Quality}
\label{sec:background:rq}

Organizations use requirements specifications in several subsequent software engineering activities.
Non-functional requirements influence a system's architecture, functional requirements determine the input and expected output of the system's features, and all requirements are ultimately translated into test cases to assert whether the developed system meets the customers' expectations.
However, quality defects in requirements specifications like missing or ambiguous requirements impede this reuse~\cite{montgomery2022empirical,wagner2019status}.

Two factors aggravate the impact of requirements quality defects on subsequent activities.
Firstly, requirements specifications are predominantly written in natural language~\cite{franch2023state} (NL).
The inherent complexity and ambiguity of NL benefits quality defects.
Secondly, the cost to remove a quality defect scales the longer it remains undetected~\cite{boehm1988understanding}.
Clarifying and rewriting an ambiguous requirement takes significantly less time than re-implementing a wrong feature built based on a misunderstood requirement.

Consequently, organizations aim to detect and remove relevant requirements quality defects as early as possible~\cite{montgomery2022empirical}.
A popular frame for this is the requirements quality factor.
A quality factor is a normative metric that maps a requirements specification to a level of quality~\cite{frattini2022live}.
For example, the quality factor \textit{passive voice} associates the use of passive voice in a requirements sentence with bad quality due to the potential omission of the subject of the sentence~\cite{rosadini2017using}.
Requirements quality research abounds with quality factors and automatic tools to detect violations against them~\cite{frattini2022live}.

\subsection{Requirements Quality Theory}
\label{sec:background:rqt}

Despite their usability, requirements quality factors suffer from significant shortcomings.
Most significantly, the majority of them lack empirical evidence for their relevance~\cite{montgomery2022empirical,frattini2022live,frattini2023requirements}, i.e., they are purely normative. 
Most publications empirically investigate the performance of a tool automatically detecting a violation against a quality factor, but only the fewest empirically investigate whether this violation does have an actual impact and is, therefore, even worth detecting and mitigating.
This undermines the practical relevance of requirements quality research, and research results are rarely adopted in practice~\cite{femmer2018requirements,franch2020practitioners}.

In response, previous research proposed a harmonized \textit{requirements quality theory} (RQT).
This theory frames requirements quality as the impact that properties of requirements specifications have on the properties of dependent activities in a given context~\cite{femmer2015activities}.
Consequently, the RQT does not consider a violation against a quality factor as harmful in itself, but only if this violation has an impact on the activities that use the requirement as an input.

\begin{figure}[hbt]
    \vspace{-.4cm}
    \centering
    \includegraphics[width=\textwidth]{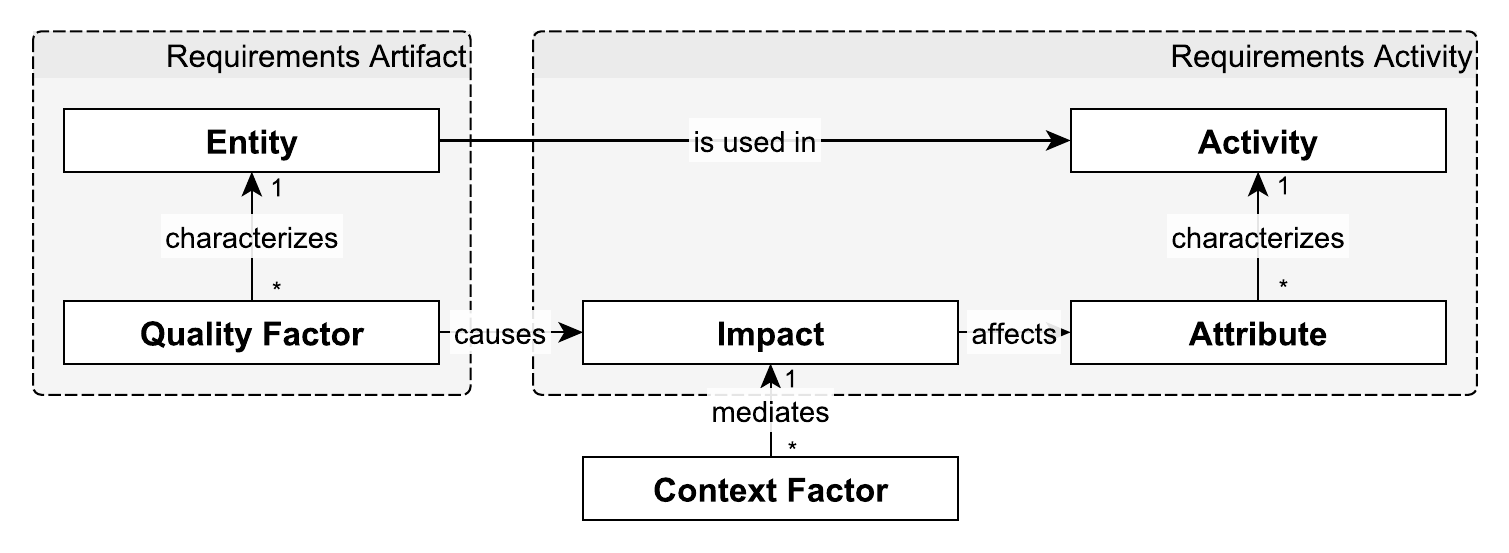}
    \caption{Groups and concepts of the reduced Requirements Quality Theory~\cite{frattini2023requirements}}
    \label{fig:rqt}
    \vspace{-.4cm}
\end{figure}

\Cref{fig:rqt} visualizes a reduced version of the RQT~\cite{frattini2023requirements}.
The RQT consists of three groups of concepts: artifact concepts, the context concept, and activity concepts.
The concept of an \textit{entity} represents all types of requirements artifacts like use cases or sentences.
\textit{Quality factors} are properties of these entities~\cite{frattini2022live}, like the length of a specification, the completeness of a use case, or the voice (active or passive) of a single sentence.
\textit{Activities} represent every software engineering process that uses a requirements entity as an input~\cite{femmer2015activities}.
This includes activities like implementing or generating test cases but also more implicit processes like understanding an entity and estimating its effort.
\textit{Attributes} are properties of these activities and include metrics like the duration or correctness of an activity.
\textit{Context factors} represent properties of the process and involved stakeholders~\cite{petersen2009context}, like the domain knowledge of involved engineers, the used process model, or the distribution of the organization.
Finally, the \textit{impact} represents the relationship between the quality factors, context factors, and attributes.

Not only does the RQT guide the framing of requirements quality, but it also enables operationalization in practice~\cite{frattini2023requirements}.
By measuring quality factors, context factors, and attributes, all input and output variables to the impact become quantified.
Once all variables are quantified, a statistical model trained on historical data can estimate the probability that a certain quality factor in a given context will affect the attribute of an activity~\cite{frattini2023requirements}.
This quantification was not attempted yet~\cite{femmer2017rapid} but advocated for in requirements quality research roadmaps~\cite{femmer2018requirements,frattini2023requirements} since it allows to (1) empirically assess and compare the criticality of quality factors, and (2) predict how a requirements specification will impact dependent activities.
This prediction model would meet the initially mentioned need of organizations to reliably detect quality defects that impact the software engineering process.

\subsection{Gap}
\label{sec:background:gap}

Requirements quality research faces two major gaps.
Firstly, requirements quality research lacks empirical evidence for the relevance of requirements quality factors~\cite{montgomery2022empirical,frattini2022live,frattini2023requirements}.
This entails the risk that requirements quality research does not focus on problems relevant to practice.

Secondly, and by extension, the RQT is difficult to operationalize without empirical evidence about the relevance of quality factors.
Previous research has already identified 206 mostly normative requirements quality factors~\cite{frattini2022live}.
Measuring all of them is not feasible, given their amount and complexity.
Empirical evidence about the relevance of quality factors will aid in prioritizing and selecting factors to consider in statistical models for impact estimation.

\section{Method}
\label{sec:method}

This study aims to address the gaps outlined in \Cref{sec:background:gap} by contributing empirical evidence to relevant factors of requirements quality.
The study addresses the following research questions:

\begin{itemize}
    \item RQ1: Which factors of requirements quality do engineers that process requirements perceive to be relevant?
    \item RQ2: Which factors of requirements quality are reported in issues?
\end{itemize}

The study contrasts two perspectives: relevant factors of requirements quality as \textit{perceived} by engineers using requirements (RQ1) and as \textit{reported} in issues (RQ2).
The study employs case study research to obtain insights on the necessary level of detail at the expense of generalizability~\cite{runeson2012case}.
The contemporary software engineering phenomenon~\cite{wohlin2021case} that is subject of the study is the impact of requirements quality defects as described through its factors.
The case study research method lends itself to the investigation since the boundary between the phenomenon and the context is unclear~\cite{wohlin2021case}.

\begin{figure}
    \centering
    \includegraphics[width=\textwidth]{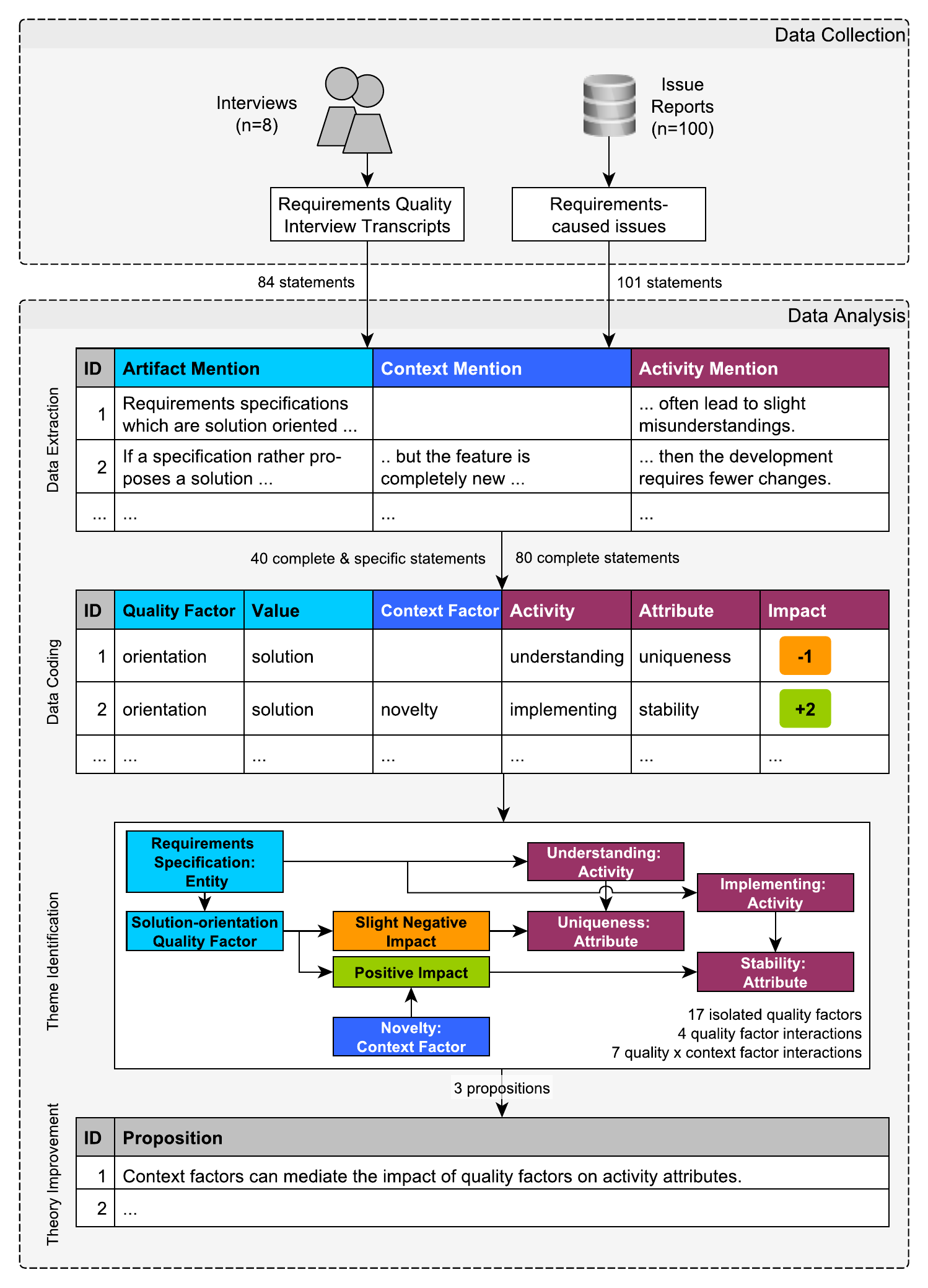}
    \caption{Overview of the data collection and analysis method}
    \label{fig:method}
\end{figure}

The method follows Runeson et al.~\cite{runeson2012case} and is reported according to the guidelines by Höst et al.~\cite{host2007checklists}.
%\Cref{sec:method:collection} describes the data collection procedure and \Cref{sec:method:analysis} explains the data analysis.
\Cref{fig:method} visualizes the steps of the process.
Verbatim examples shown in the figure and throughout this section (in quotation marks) are artificial as the raw data is confidential.

\subsection{Data Collection}
\label{sec:method:collection}

Ericsson, the case company providing the data, is a large, globally distributed software development organization.
Ericsson follows an agile development approach but completes the \textit{requirements specification} for each new feature or change request before committing to subsequent phases like design, implementation, and testing.
These requirements specifications use unconstrained natural language to specify one or more requirements related to a specific feature or change request.
The study evaluates two different types of data to triangulate and strengthen the results~\cite{wohlin2021case}. %,bratthall2002can}.
The upper section of \Cref{fig:method} visualizes the two data collection approaches.

\subsubsection{Interview Data}

To understand the factors that engineers processing requirements perceive to be relevant, interviews were conducted (RQ1).
%Interviews were conducted to understand which factors of requirements quality stakeholders perceive to be relevant (RQ1).
%This data classifies as \textit{first degree} data given the involvement of a researcher in the collection process~\cite{lethbridge2005studying}.
A contact at the case company provided a sample of eight software engineers responsible for processing the requirements specification and developing a solution specification.
The interview participants had an average of 3.5 years of experience in their role, 7.5 years with the organization, and 15.3 years as software engineers.

A protocol aided in conducting the semi-structured interviews based on previous requirements quality research.
In particular, the eleven themes of requirements quality by Montgomery et al.~\cite{montgomery2022empirical} served as sections of the interview.
These themes include, for example, ambiguity, completeness, and correctness.
For each theme, the interview participants were asked whether they experienced any issues of this type when processing requirements.
If yes, they were prompted to elaborate on the issue using the RQT as a frame~\cite{frattini2023requirements}.
Otherwise, the interview moved to the next theme.
A slide deck introduced and visualized the RQT to the interview participants to aid the conversation.
The author of this paper conducted all eight interviews, each taking up to one hour.

The recorded interview sessions were automatically transcribed using Descript\footnote{\url{https://www.descript.com/}}.
Afterwards, the author manually checked all transcripts and ensured that the automatic transcription matched the recording.
The replication package contains all supplementary material, including the interview guidelines.
The transcripts contain confidential information and cannot be shared.

\subsubsection{Issue Data}

To understand which factors of requirements quality cause a reported impact, issues from an issue tracker were analyzed (RQ2).
%Issues from an issue tracker were analyzed to understand which factors of requirements quality cause a reported impact (RQ2).
%The collected data counts as \textit{third degree} data~\cite{lethbridge2005studying} given these artifacts already existed prior to the study~\cite{runeson2012case}.
The contact at the case company provided access to Ericsson's database of issue reports.
This database contains issues raised both during the internal development process and from external customers.
Every issue denotes the development phases in which it was detected and in which the root cause of this issue has been introduced.
Domain experts procure and document the latter information.
For this study, the available issues were filtered to obtain only those that have been introduced during the requirements engineering phase, resulting in 100 issue reports from January 2021 until September 2023.

\subsection{Data Analysis}
\label{sec:method:analysis}

To analyze the large body of textual data, thematic synthesis as proposed by Cruzes and Dyb\aa ~\cite{cruzes2011recommended} was employed and reported according to their guideline. 
The lower section of \Cref{fig:method} visualizes the data analysis steps.

\subsubsection{Data extraction}

Extracting relevant data from the textual corpus comprised the first step. 
Each defect perceived by an interview participant constitutes one statement about requirements quality. % containing up to three mentions.
The eight interview transcripts produced 84 statements about perceived requirements quality.
The 100 issues contained 101 statements about reported requirements quality.
The three groups of the RQT (artifact, context, and activity) provided a frame for the data extraction.
For each of the three groups, all relevant mentions from a statement were extracted.
For example, the statement ``If a specification rather proposes a solution, but the feature is completely new, then the development requires fewer changes.'' contains one mention of each group:
``a specification rather proposes a solution'' describes a property of the artifact, ``the feature is completely new'' describes the context, and ``the development requires fewer changes'' represents the activity impacted by the quality and context factor.

\subsubsection{Data coding}

Each mention received codes for all the concepts contained within the respective RQT group.
An artifact mention can contain a number of quality factors with a value associated with each of them.
A context mention can contain a number of context factors.
An activity mention can contain a number of activities, each associated with one attribute and an impact value.
All codes and coding instructions were documented in extensive coding guidelines.
%The development of an extensive coding guideline preceded this step.

\begin{figure}
    \centering
    \includegraphics[width=\textwidth]{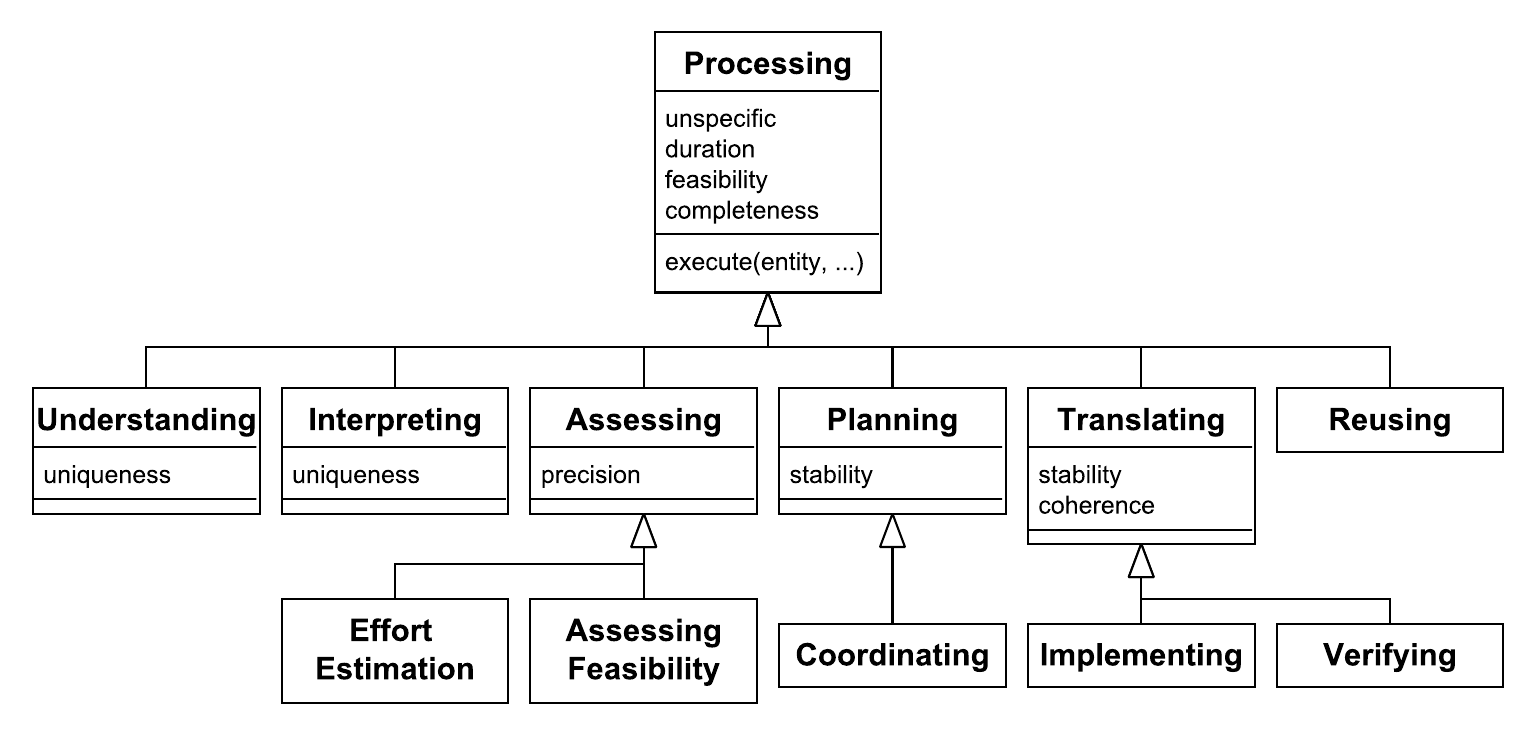}
    \caption{Model of activities (as classes) and attributes (as their attributes)}
    \label{fig:activities}
\end{figure}

Coding the artifact and context mentions followed a deductive approach~\cite{miles1994qualitative}:
quality factors~\cite{frattini2022live,montgomery2022empirical} and context factors~\cite{petersen2009context} identified during previous research constituted the available codes.
%because existing literature on quality factors~\cite{frattini2022live,montgomery2022empirical} and context factors~\cite{petersen2009context} allowed to develop codes prior to the coding process. %provides sufficient background for preconceived codes.
Coding the activity mentions followed an inductive approach~\cite{strauss1990basics} since literature is still lacking in this regard~\cite{femmer2018requirements,frattini2023requirements}.
\Cref{fig:activities} visualizes the activity and attribute codes generated inductively from the data.
Descriptions of all activities and attributes can be found in the coding guideline contained in the replication package.

The artifact mention ``if a specification rather proposes a solution'' received the quality factor code \textit{orientation} and the value \textit{solution}, since it describes a \textit{solution-oriented} requirements specification.
The context factor mention ``but the feature is completely new'' was coded \textit{novelty}.
Finally, the activity mention ``then the development requires less changes'' received the following codes:
the \textit{stability} attribute of the \textit{implementing} activity experiences a \textit{positive} impact (\impactgood).
We coded the strength of the impact on a discrete seven-point scale. 
Positive (\impactgood), negative (\impactbad), and no impact (\impactnone) were used as the default codes depending on the direction of the impact.
Particularly strong (\impactverygood~and \impactverybad) or weak (\impactslightlygood~and \impactslightlybad) codes were only used when explicitly mentioned by the interview participant, e.g., if an impact was ``very critical'' or just ``slight.''

Independent coders were involved to ensure the reliability of the subjective coding task.
For the coding of the interview transcripts, a senior researcher from the same department as the author, who is also familiar with the case company, independently coded ten randomly selected statements using the coding guideline.
The inter-rater agreement of codes achieved a percentage score of $83.8\%$ and a Cohen's Kappa of $71.8\%$. %~\cite{cohen1960coefficient} of $71.8\%$. 
However, Cohen's Kappa is known to be unreliable for uneven marginal distributions~\cite{feng2015mistakes}.
Still, the more robust S-Score~\cite{bennett1954communications} resulted in an agreement of $82.3\%$.
All scores represent a substantial agreement and support the reliability of the coding process.

For the coding of the issue reports, a senior engineer from the case company was involved.
In a session between the author and the involved engineer, questionable codes were reviewed and adjusted.
This process confirmed the applicability of the chosen codes as well as that 21 of the selected issues were not caused by requirements quality defects but unrelated circumstances.
These 21 statements were not considered in future phases of the data analysis.

\subsubsection{Theme identification}

In this step, individual codes are aggregated into ``more meaningful and parsimonious units''~\cite{cruzes2011recommended}.
The theme identification imposes two further conditions on the coded statements:
(C1) A statement must contain at least one quality factor and at least one activity, and
(C2) a statement must not contain any unspecific activity (activity code: \textit{processing}, root of \Cref{fig:activities}) or an unspecific attribute (attribute code: \textit{unspecific}).
This rules out 44 vague interview statements of the form ``\textit{quality factor} is bad'' (C1, missing activity) or ``\textit{quality factor} is bad for implementing'' (C2, unspecific attribute).

The final analysis considered 40 interview and 80 issue statements.
The amount of information per mention differed between the two data sources.
While the interview encouraged participants to elaborate on all concepts of the RQT, the issue data lacked the same level of control~\cite{runeson2012case}.
Issue statements always contained the activity concepts since they explicitly report the effect of an issue, but the level of information on the root cause in requirements engineering was often limited, and the issues did not contain any information about the context.

%We analyze the more granular and complete interview statements in three groups
The analysis of the more granular and complete interview statements splits the data into three groups: statements containing a single quality factor and no context factors, statements containing multiple quality factors but no context factors, and statements containing context factors.
Within each group, statements about the same quality factors are aggregated to collect all impacted activities and attributes.
Since the issue statements contained codes on higher levels and no context factors, they were aggregated into one matrix showing the distribution of quality factors impacting activity attributes.

\subsubsection{Model creation}

The final inferential step of the guideline by Cruzes et al. is the description of higher-order themes, a taxonomy, model, or a theory~\cite{cruzes2011recommended}.
Because this study is already grounded in the theoretical foundation provided by the RQT~\cite{frattini2023requirements}, but none of the encountered data challenged this theory, developing a new model or theory was not deemed constructive.
Instead, this study evolves the existing RQT by deriving propositions from the identified themes.
These propositions enrich the theory with empirical insights and contribute falsifiable hypotheses for further research.

\subsubsection{Trustworthiness Assessment}

The final overall step of the guideline by Cruzes et al. is to assess the trustworthiness of the synthesis~\cite{cruzes2011recommended}.
As these concerns align with threats to validity, they are addressed in \Cref{sec:discussion:threats}.

\section{Results}
\label{sec:result}

%\Cref{sec:results:interview} presents the identified themes from the interview data, and %\Cref{sec:results:issues} the themes from the issue reports.
%\Cref{sec:results:propositions} lists the propositions derived from both sets of themes.

\subsection{Interview data}
\label{sec:results:interview}

\subsubsection{Impact of single Quality Factors}

The interview data contains information about 17 unique quality factors with an impact on at least one subsequent activity.
For brevity, \Cref{tab:impact:single} lists only the four quality factors that were contained in at least two statements. 
Each cell of the \textit{impact} column in \Cref{tab:impact:single} lists the perceived direction and strength of the impact that the quality factor has on the attribute of the activity in this cell.
For example, two statements of the interview data described a negative impact of solution-oriented requirements on a unique understanding of that requirement.
One statement stressed that this impact is major (\impactverybad), the other one did not (\impactbad).
The replication package contains the remaining quality factors and their impact\footnote{Available at \url{https://github.com/JulianFrattini/rqi-relf/blob/main/src/analytics/results.md}}.

\begin{comment}
\begin{table}[hbt!]
    \centering
    \caption{Perceived impact of single quality factors on subsequent activities}
    \label{tab:impact:single}
    \begin{tabular}{l|cc|c|cc|c|cc|c} 
        \toprule
        \multicolumn{1}{r|}{\textbf{Activity} } & \multicolumn{2}{c|}{\rotatebox[origin=c]{270}{Understanding}} & \rotatebox[origin=c]{270}{Interpreting} & \multicolumn{2}{c|}{\rotatebox[origin=c]{270}{Verifying}} & \rotatebox[origin=c]{270}{Effort Estimation} & \multicolumn{2}{c|}{\rotatebox[origin=c]{270}{Translating}} & \rotatebox[origin=c]{270}{Planning} \\ 
        \multicolumn{1}{r|}{\textbf{Attribute} } & \rotatebox[origin=c]{270}{Unique.} & \rotatebox[origin=c]{270}{Duration} & \rotatebox[origin=c]{270}{Uniqueness} & \rotatebox[origin=c]{270}{Complete.} & \rotatebox[origin=c]{270}{Duration} & \rotatebox[origin=c]{270}{Traceability} & \rotatebox[origin=c]{270}{Duration} & \rotatebox[origin=c]{270}{Stability} & \rotatebox[origin=c]{270}{Stability} \\ \midrule
        \textbf{Quality Factor} & & & & & & & & & \\ \midrule
        Solution-orientation & \impactverybad~\impactbad & & & \impactbad~\impactbad & & \impactgood & & \impactgood & \impactgood \\
        Non-atomic & & & & & & & \impactbad~\impactbad & & \impactbad \\
        Non-concise & \impactslightlybad & \impactbad & &  & & & & & \\
        Too dense & & \impactbad & \impactbad & & \impactbad & & & \\
        \bottomrule
    \end{tabular}
\end{table}
\end{comment}

\begin{table}[hbt!]
    \centering
    \caption{Perceived impact of single quality factors on subsequent activities}
    \label{tab:impact:single}
    \begin{tabular}{l|ll|c}
    \toprule
        \textbf{Quality Factor} & \textbf{Activity} & \textbf{Attribute} & \textbf{Impact} \\
        \midrule
        \multirow{6}{*}{Solution-orientation} & Understanding & Uniqueness & \impactverybad~\impactbad \\
        & Verifying & Completeness & \impactbad~\impactbad \\
        & Effort Estimation & Traceability & \impactgood \\
        & Translating & Stability & \impactgood \\
        & Feasibility Assessment & Precision & \impactgood \\
        & Planning & Stability & \impactgood \\ 
        \hline
        \multirow{2}{*}{Non-atomic} & Translating & Duration & \impactbad~\impactbad \\
        & Planning & Stability & \impactbad \\
        \hline
        \multirow{2}{*}{Non-concise} & \multirow{2}{*}{Understanding} & Uniqueness & \impactslightlybad \\
        & & Duration & \impactbad \\
        \hline
        \multirow{3}{*}{Too dense} & Understanding & Duration & \impactbad \\
        & Interpreting & Uniqueness & \impactbad \\
        & Verifying & Duration & \impactbad \\
        \bottomrule
    \end{tabular}
    \vspace{-.4cm}
\end{table}

The table shows that the interview participants perceived the four most often mentioned quality factors to impact a variety of activities and their attributes.
The most frequently mentioned quality factor is \textit{solution-orientation}, i.e., requirements that impose on the solution space rather than elaborating on the problem space~\cite{fernandez2012field}.
This quality factor is perceived to cause misunderstandings (i.e., causing the \textit{understanding} activity to be not \textit{unique}) and lack of coverage when deriving test cases (i.e., causing the \textit{verifying} activity to be not \textit{complete}).
The table also shows that some quality factors have a mixed impact on different activities.
For example, a solution-oriented requirements specification is also perceived to aid effort estimation, translating, and planning.

\subsubsection{Interaction of multiple Quality Factors}

The interview data contains information about four unique interactions between quality factors.
\Cref{tab:impact:multiple} lists the four interaction effects.
The interview participants reported that redundant requirements that were also not connected through horizontal trace links (i.e., links between requirements) caused incoherent implementations.
Furthermore, non-functional requirements were reported to be susceptible to ambiguous understanding when providing to little details.
Additionally, requirements specifications that were yet immature but also already committed to were quicker to implement due to the applied time pressure (\impactgood), but implementation became much less feasible (\impactverybad).
Finally, the precision of feasibility assessment suffered from jargonic and dense requirements.

\begin{table}[hbt!]
    \vspace{-.6cm}
    \centering
    \caption{Interaction between two quality factors}
    \label{tab:impact:multiple}
    \begin{tabular}{p{3cm}p{3cm}|ll|c}
        \toprule
        \textbf{Quality Factor 1} & \textbf{Quality Factor 2} & \textbf{Activity} & \textbf{Attribute} & \textbf{Impact}\\ \midrule
        Redundancy & Missing horizontal trace links & Implementing & Coherence & \impactbad \\ \hline
        Too little details & Non-functional requirement & Understanding & Uniqueness & \impactbad \\ \hline
        \multirow{2}{*}{Immature} & \multirow{2}{*}{Committed} & \multirow{2}{*}{Implementing} & Duration & \impactgood \\
        & & & Feasibility & \impactverybad \\ \hline
        Jargonic & Too dense & Assessing Feasibility & Precision & \impactbad \\ 
        \bottomrule
    \end{tabular}
    \vspace{-.4cm}
\end{table}

\subsubsection{Interaction with Context Factors}

\begin{figure}
    \centering
    \includegraphics[width=\textwidth]{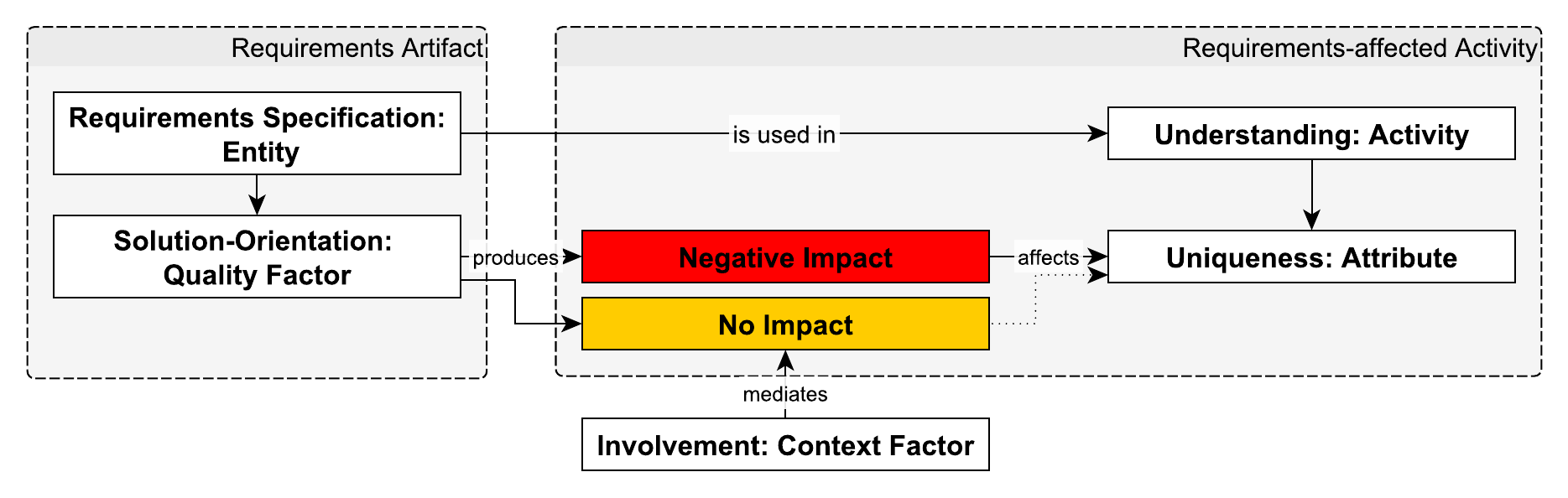}
    \caption{Interaction effect between \textit{solution-orientation} and \textit{involvement} on the \textit{understanding} activity}
    \label{fig:context:involvement}
    \vspace{-1.2cm}
\end{figure}

The interview data contains information about seven unique interactions between quality factors and context factors.
The two most prominent prominently perceived interaction effects involve the quality factor \textit{solution-orientation} and \textit{density}.
\Cref{fig:context:involvement} visualizes one statement describing the interaction between the quality factor \textit{solution-orientation} and context factor \textit{involvement} on the \textit{uniqueness} attribute of the \textit{understanding} activity.
If the stakeholder responsible for processing the requirement was also involved in writing it, the impact on the understandability is mitigated.
The remaining interaction effects are detailed in our replication package but follow a similar pattern:
Context factors like \textit{involvement}, \textit{experience}, and \textit{supplementary communication} can mitigate the negative impact of quality factors.
Additionally, solution-oriented requirements exhibit an even stronger positive impact on several activities like feasibility assessment and effort estimation when the context of the requirement is \textit{new}.

\subsection{Issue reports}
\label{sec:results:issues}

The issue data only allows to infer general relationships between quality factors and activities.
\Cref{tab:issues} lists the number of statements per constellation of quality factor, activity, and attribute.
The most prominent impact of requirements quality that results in a reported issue is \textit{completeness}, i.e., caused by a missing requirement.
This results mostly in \textit{incorrect} (i.e., bugs) or \textit{incomplete} (i.e., missing features) implementations.
In rare cases, the \textit{understanding}, \textit{interpreting}, and \textit{verifying} activity are reported to be impacted.
Behind completeness, the most often reported impact is \textit{consistency} and \textit{ambiguity}. 

\begin{table}[hbt!]
    \centering
    \caption{Requirements impact as recorded in  the issue reports (Comp. = Completeness, Corr. = Correctness, Cons. = Consistency)}
    \label{tab:issues}
    \begin{tabular}{l|c|c|ccc|cc}
        \toprule
        \multicolumn{1}{r|}{\textbf{Activity} } & Understanding & Interpreting & \multicolumn{3}{c|}{Implementing} & \multicolumn{2}{c}{Verifying } \\
        \multicolumn{1}{r|}{\textbf{Attribute} } & Unique & Unique & Comp. & Corr. & Cons. & Coverage & Feasible \\ \midrule
        \textbf{Quality Factor} & & & & & & & \\ \midrule
        Completeness & 1 & 0 & 13 & 34 & 3 & 1 & 1 \\
        Consistency & 1 & 0 & 0 & 6 & 2 & 2 & 0 \\
        Ambiguity & 1 & 2 & 2 & 1 & 2 & 0 & 0 \\
        Correctness & 1 & 0 & 0 & 2 & 1 & 0 & 0 \\
        Feasibility & 0 & 0 & 0 & 1 & 0 & 1 & 0 \\
        Relevance & 0 & 0 & 1 & 0 & 0 & 0 & 0 \\
        \bottomrule
    \end{tabular}
\end{table}

\subsection{Propositions}
\label{sec:results:propositions}

The triangulation of interview and issue analysis results allows to derive the following propositions enhancing the RQT.

\subsubsection{Relevant quality factors}

The set of quality factors perceived and reported to be relevant in the case company is very limited.
Among the perceived quality factors, only solution-orientation, lack of atomicity, lack of conciseness, and density received support in at least two statements.
Among the reported quality factors, lack of completeness (i.e., missing requirements) stands out as the primary cause of issues in the down-stream development process.

\subsubsection{Mixed Impact}

The analysis of the interview data shows that quality factors can have mixed impact on different activities.
For example, while a solution-oriented requirement might negatively impact the activities of understanding and verifying it, planning and translating activities become more stable.

\subsubsection{Interactions matter}

The analysis of the interview data shows that interactions between quality factors but also between quality and context factors have a significant effect.
In particular, quality factors like the novelty of a feature, the experience of the involved engineers, and supplementary communication mediate the effect of quality factors as shown in \Cref{fig:context:involvement}.

\section{Discussion}
\label{sec:discussion}

%\Cref{sec:discussion:implications} discusses the implications of the results presented in %\Cref{sec:result}, and in \Cref{sec:discussion:threats} the threats to their validity.

\subsection{Implications}
\label{sec:discussion:implications}

\subsubsection{Contribution to Research}

The empirical evidence both strengthens existing requirements engineering theories and guides further advances in requirements quality research.
The results strengthen the RQT~\cite{frattini2023requirements,femmer2015activities} in that artifact properties impacting activity properties constitutes requirements quality.
Moreover, the results confirm that one quality factor may have different impacts on different activities~\cite{femmer2015activities}.
Incautiously removing a quality factor from an entity due to the negative impact on one activity may, therefore, also mitigate its positive impact on other activities.
The results further strengthen the \textit{Naming the Pain in Requirements Engineering} (NaPiRE) initiative~\cite{mendez2017naming} by contributing more granular evidence to the problems of RE relevant to practice.
The results agree with the conclusion of NaPiRE that missing requirements are among the most impactful quality defects.
Similarly, the results of both data sources of the study agree with previous studies that the effect of ambiguity is less relevant in practice than often assumed~\cite{philippo2013requirement,de2010ambiguity}.
Finally, the results support the advocacy for context-sensitive research in empirical software engineering~\cite{briand2017case} by emphasizing context factors as mediators of the impact of artifacts on activities.
The results of this study guide further research advances through the approach of identifying relevant factors of requirements quality.
The study reduces the vast space of several hundred potential quality factors~\cite{frattini2022live} down to about 30 that are relevant to the specific context of an organization.

\subsubsection{Impact on Practice}

The operationalization of the RQT in practice becomes feasible due to the reduction of variables to measure.
The results steer the next step of research with the case company toward detecting solution-oriented, non-atomic, non-concise, dense, and incomplete requirements.
Additionally, effort will be focused on measuring the relevant context factors of involvement, experience, novelty, and supplementary communication.
These measurements enable the impact estimation of requirements quality on dependent activities and advance requirements quality research roadmaps~\cite{femmer2018requirements,frattini2023requirements}.

\subsubsection{Limitations}

One gap to overcome is the lack of an overview of requirements-dependent activities and their measurable attributes, as also outlined in previous research roadmaps~\cite{femmer2018requirements,frattini2023requirements}.
These attributes constitute the dependent variable in the impact estimation implied by the RQT~\cite{frattini2023requirements}.
The interview data contains 44 statements with either an unclear activity or attribute impacted by a quality factor.
Developing a model of requirements-dependent activities and their attributes is a necessary next step to achieve operationalization of the RQT~\cite{frattini2023requirements}.

\subsection{Threats to Validity}
\label{sec:discussion:threats}

This section discusses threats to validity according to Runesson et al.~\cite{runeson2012case} and Wohlin et al.~\cite{wohlin2012experimentation} and, additionally, addresses concerns of trustworthiness of the thematic synthesis~\cite{cruzes2011recommended}.

%\subsubsection{Conclusion validity}
% Threats to the conclusion validity are concerned with issues that affect the ability to draw the correct conclusion about relations between the treatment and the outcome of an experiment.
Regarding conclusion validity, the \textit{reliability of measures} is a prevalent threat given the subjective coding process but has been minimized by involving independent raters and calculating inter-rater agreement where applicable.
Similarly, the lack of control over the issue data questions their reliability.
Involving a senior engineer from the case company mitigated this issue, providing an adequate \textit{confirmability} of the data.
%We did not control for \textit{dependability} of the data, i.e., we cannot state whether the the data is stable or changes over time.

%\subsubsection{Internal validity}
% Threats to internal validity are influences that can affect the independent variable with respect to causality, without the researcher’s knowledge.
Regarding internal validity, the interview data suffers from \textit{selection} bias.
The interview participants were sampled by the industry contact.
However, the participants show a wide variation of background and experience, which leads to assume that this threat is minimal.

%\subsubsection{Construct Validity}
% Construct validity concerns generalizing the result of the experiment to the concept or theory behind the experiment.
Regarding construct validity, the study suffers from \textit{inadequate preoperational explication of constructs}, i.e., immaturity of some of the concepts of the theoretical framework (the RQT).
In particular, the activity group within the RQT (activities and attributes) is insufficiently explored in requirements quality research~\cite{femmer2018requirements,frattini2023requirements}.
As a consequence, several interview statements failed to specify the activity and attribute impacted by requirements quality.
The threat was minimized by excluding this data from the analysis.

%\subsubsection{External Validity} 
% Threats to external validity are conditions that limit our ability to generalize the results of our experiment to industrial practice.
Regarding external validity, the inference of this study is not generalizable or transferable~\cite{cruzes2011recommended} by design of the case study method.
Additional research replicating this study in other companies are necessary to generalize the results.

\section{Conclusion}
\label{sec:conclusion}

This case study demonstrates the application of the requirements quality theory (RQT) to identify relevant factors of requirements quality.
By analyzing both interview and issue report data, we identified 17 relevant quality as well as 11 interaction effects among them and with context factors.
The study contributes empirical evidence to the relevance of these factors and their effects in the case company.
The study emphasizes that (1) some requirements quality factors are more relevant in practice than others, (2) they may have a simultaneous negative and positive impact on different activities, and (3) context factors mediate their impact.
This research advances requirements quality research by advancing existing research roadmaps~\cite{femmer2018requirements} toward a quantified impact estimation of requirements quality in practice~\cite{frattini2023requirements}.

%\section*{Acknowledgements}

%This work was supported by the KKS foundation through the S.E.R.T. Research Profile project at Blekinge Institute of Technology.
%I thank Michael Unterkalmsteiner for independently performing the coding task.
%Furthermore, I owe great thanks to Parisa Yousefi, Charlotte Ljungman, and Fabiano Sato from Ericsson for their continuous support that made this work possible in the first place.

\bibliographystyle{splncs04}
\bibliography{references}

\begin{thebibliography}{10}
\providecommand{\url}[1]{\texttt{#1}}
\providecommand{\urlprefix}{URL }
\providecommand{\doi}[1]{https://doi.org/#1}

\bibitem{bennett1954communications}
Bennett, E.M., Alpert, R., Goldstein, A.: Communications through limited-response questioning. Public Opinion Quarterly  \textbf{18}(3),  303--308 (1954)

\bibitem{boehm1988understanding}
Boehm, B.W., Papaccio, P.N.: Understanding and controlling software costs. IEEE transactions on software engineering  \textbf{14}(10),  1462--1477 (1988)

\bibitem{briand2017case}
Briand, L., Bianculli, D., Nejati, S., Pastore, F., Sabetzadeh, M.: The case for context-driven software engineering research: generalizability is overrated. IEEE Software  \textbf{34}(5),  72--75 (2017)

\bibitem{de2010ambiguity}
de~Bruijn, F., Dekkers, H.L.: Ambiguity in natural language software requirements: A case study. In: Requirements Engineering: Foundation for Software Quality: 16th International Working Conference, REFSQ 2010, Essen, Germany, June 30--July 2, 2010. Proceedings 16. pp. 233--247. Springer (2010)

\bibitem{cruzes2011recommended}
Cruzes, D.S., Dyba, T.: Recommended steps for thematic synthesis in software engineering. In: 2011 international symposium on empirical software engineering and measurement. pp. 275--284. IEEE (2011)

\bibitem{femmer2018requirements}
Femmer, H.: Requirements quality defect detection with the qualicen requirements scout. In: REFSQ Workshops (2018)

\bibitem{femmer2017rapid}
Femmer, H., Fern{\'a}ndez, D.M., Wagner, S., Eder, S.: Rapid quality assurance with requirements smells. Journal of Systems and Software  \textbf{123},  190--213 (2017)

\bibitem{femmer2015activities}
Femmer, H., Mund, J., Fern{\'a}ndez, D.M.: It's the activities, stupid! a new perspective on re quality. In: 2015 IEEE/ACM 2nd International Workshop on Requirements Engineering and Testing. pp. 13--19. IEEE (2015)

\bibitem{feng2015mistakes}
Feng, G.C.: Mistakes and how to avoid mistakes in using intercoder reliability indices. Methodology: European Journal of Research Methods for the Behavioral and Social Sciences  \textbf{11}(1), ~13 (2015)

\bibitem{fernandez2012field}
Fernandez, D.M., Wagner, S., Lochmann, K., Baumann, A., de~Carne, H.: Field study on requirements engineering: Investigation of artefacts, project parameters, and execution strategies. Information and Software Technology  \textbf{54}(2),  162--178 (2012)

\bibitem{franch2020practitioners}
Franch, X., Mendez, D., Vogelsang, A., Heldal, R., Knauss, E., Oriol, M., Travassos, G., Carver, J.C., Zimmermann, T.: How do practitioners perceive the relevance of requirements engineering research? IEEE Transactions on Software Engineering  (2020)

\bibitem{franch2023state}
Franch, X., Palomares, C., Quer, C., Chatzipetrou, P., Gorschek, T.: The state-of-practice in requirements specification: an extended interview study at 12 companies. Requirements Engineering pp. 1--33 (2023)

\bibitem{frattini2023requirements}
Frattini, J., Montgomery, L., Fischbach, J., Mendez, D., Fucci, D., Unterkalmsteiner, M.: Requirements quality research: a harmonized theory, evaluation, and roadmap. Requirements engineering  (2023)

\bibitem{frattini2022live}
Frattini, J., Montgomery, L., Fischbach, J., Unterkalmsteiner, M., Mendez, D., Fucci, D.: A live extensible ontology of quality factors for textual requirements. In: 2022 IEEE 30th International Requirements Engineering Conference (RE). pp. 274--280. IEEE (2022)

\bibitem{host2007checklists}
Host, M., Runeson, P.: Checklists for software engineering case study research. In: First international symposium on empirical software engineering and measurement (ESEM 2007). pp. 479--481. IEEE (2007)

\bibitem{mendez2017naming}
M{\'e}ndez, D., Wagner, S., Kalinowski, M., Felderer, M., Mafra, P., Vetr{\`o}, A., Conte, T., Christiansson, M.T., Greer, D., Lassenius, C., et~al.: Naming the pain in requirements engineering: Contemporary problems, causes, and effects in practice. Empirical software engineering  \textbf{22},  2298--2338 (2017)

\bibitem{miles1994qualitative}
Miles, M.B., Huberman, A.M.: Qualitative data analysis: An expanded sourcebook. sage (1994)

\bibitem{montgomery2022empirical}
Montgomery, L., Fucci, D., Bouraffa, A., Scholz, L., Maalej, W.: Empirical research on requirements quality: a systematic mapping study. Requirements Engineering  \textbf{27}(2),  183--209 (2022)

\bibitem{petersen2009context}
Petersen, K., Wohlin, C.: Context in industrial software engineering research. In: 2009 3rd International Symposium on Empirical Software Engineering and Measurement. pp. 401--404. IEEE (2009)

\bibitem{philippo2013requirement}
Philippo, E.J., Heijstek, W., Kruiswijk, B., Chaudron, M.R., Berry, D.M.: Requirement ambiguity not as important as expected—results of an empirical evaluation. In: Requirements Engineering: Foundation for Software Quality: 19th International Working Conference, REFSQ 2013, Essen, Germany, April 8-11, 2013. Proceedings 19. pp. 65--79. Springer (2013)

\bibitem{rosadini2017using}
Rosadini, B., Ferrari, A., Gori, G., Fantechi, A., Gnesi, S., Trotta, I., Bacherini, S.: Using nlp to detect requirements defects: An industrial experience in the railway domain. In: Requirements Engineering: Foundation for Software Quality: 23rd International Working Conference, REFSQ 2017, Essen, Germany, February 27--March 2, 2017, Proceedings 23. pp. 344--360. Springer (2017)

\bibitem{runeson2012case}
Runeson, P., Host, M., Rainer, A., Regnell, B.: Case study research in software engineering: Guidelines and examples. John Wiley \& Sons (2012)

\bibitem{strauss1990basics}
Strauss, A., Corbin, J.: Basics of qualitative research. Sage publications (1990)

\bibitem{wagner2019status}
Wagner, S., Fern{\'a}ndez, D.M., Felderer, M., Vetr{\`o}, A., Kalinowski, M., Wieringa, R., Pfahl, D., Conte, T., Christiansson, M.T., Greer, D., et~al.: Status quo in requirements engineering: A theory and a global family of surveys. ACM Transactions on Software Engineering and Methodology (TOSEM)  \textbf{28}(2),  1--48 (2019)

\bibitem{wohlin2021case}
Wohlin, C.: Case study research in software engineering—it is a case, and it is a study, but is it a case study? Information and Software Technology  \textbf{133},  106514 (2021)

\bibitem{wohlin2012experimentation}
Wohlin, C., Runeson, P., H{\"o}st, M., Ohlsson, M.C., Regnell, B., Wessl{\'e}n, A.: Experimentation in software engineering. Springer Science \& Business Media (2012)

\end{thebibliography}

\end{document}